\documentclass[sigconf]{acmart}
\AtBeginDocument{%
  }

\setcopyright{acmlicensed}
\copyrightyear{2025}
\acmYear{2025}
\acmDOI{XXXXXXX.XXXXXXX}
\acmConference[Conference acronym 'XX]{Make sure to enter the correct
  conference title from your rights confirmation email}{June 03--05,
  2018}{Woodstock, NY}
\acmISBN{978-1-4503-XXXX-X/2018/06}

\usepackage{todonotes}

\usepackage{booktabs}

\usepackage{algorithm}
\usepackage[indLines=false]{algpseudocodex}
\newcommand{\journal}[1]{}
\usepackage{mdframed}
\usepackage{pifont}
\usepackage{booktabs}
\usepackage{wrapfig}
\usepackage{url}
\usepackage{graphicx}
\usepackage{tabularx}
\usepackage{colortbl}
\usepackage{enumitem}
\usepackage{enumitem}
\usepackage{todonotes}
\usepackage{subfig}
\usepackage[export]{adjustbox}
\usepackage[normalem]{ulem}
\usepackage{balance}

\usepackage{listings}
\definecolor{lstback}{rgb}{0.96, 0.96, 0.96}
\usepackage{titlesec}
\titlespacing*{\section}
{0pt}{2.5ex plus 1ex minus .2ex}{1.4ex plus .2ex}
\titlespacing*{\subsection}
{0pt}{1.8ex plus 1ex minus .2ex}{0.9ex plus .2ex}
\titlespacing*{\subsubsection}
{0pt}{1.5ex plus 1ex minus .2ex}{0.6ex plus .2ex}

\usepackage{enumitem}
\setlist{nolistsep}

\renewcommand{\tt}[1]{\texttt{#1}}

\usepackage [autostyle, english = american]{csquotes}
\MakeOuterQuote{"}

\newcommand{\rqone}{Is it possible to track parse tree nodes through time even across unparseable states?}
\newcommand{\rqtwo}{Can the findings of \citet{brown2024writing} be replicated using automated methods?}
\newcommand{\rqthree}{What programming constructs have higher node deletion rate?}
\newcommand{\rqfour}{How much do students jump around in their code while programming?}

\begin{document}

\title{Parse Tree Tracking Through Time for Programming Process Analysis at Scale}

\author{Matt Rau}
\affiliation{%
  \institution{Utah State University}
  \city{Logan}
  \country{Utah}}
\email{matt.rau@usu.edu}

\author{Chris Brown}
\affiliation{%
  \institution{Utah State University}
  \city{Logan}
  \country{Utah}}
\email{christopher.r.brown.46@gmail.com}

\author{John Edwards}
\affiliation{%
  \institution{Utah State University}
  \city{Logan}
  \country{Utah}}
\email{john.edwards@usu.edu}




\renewcommand{\shortauthors}{Trovato et al.}


\begin{abstract}
\textbf{Background and Context:} Programming process data can be utilized to understand the processes students use to write computer programming assignments. Keystroke- and line-level event logs have been used in the past in various ways, primarily in high-level descriptive statistics (e.g., timings, character deletion rate, etc). Analysis of behavior in context (e.g., how much time students spend working on loops) has been cumbersome because of our inability to automatically track high-level code representations, such as abstract syntax trees, through time and unparseable states.

\noindent\textbf{Objective:} Our study has two goals. The first is to design the first algorithm that tracks parse tree nodes through time. Second, we utilize this algorithm to perform a partial replication study of prior work that used manual tracking of code representations, as well as other novel analyses of student programming behavior that can now be done at scale.
    
\noindent\textbf{Method:} We use two algorithms presented in this paper to track parse tree nodes through time and construct tree representations for unparseable code states. We apply these algorithms to a public keystroke data from student coursework in a 2021 CS1 course and conduct analysis on the resulting parse trees.
    
\noindent\textbf{Findings:} We discover newly observable statistics at scale, including that code is deleted at similar rates inside and outside of conditionals and loops, a third of commented out code is eventually restored, and that  frequency with which students jump around in their code may not be indicative of struggle.

\noindent\textbf{Implications:} The ability to track parse trees through time opens the door to understanding new dimensions of student programming, such as best practices of structural development of code over time, quantitative measurement of what syntactic constructs students struggle most with, refactoring behavior, and attention shifting within the code.
\end{abstract}

\ccsdesc[500]{Social and professional topics~CS1}

\keywords{CS1, keystrokes, programming process, parse trees}


\maketitle

\section{Introduction}
In this paper, we present a framework for analyzing the evolution of students' code as they write programming assignments. Most works up until now have focused on high-level statistics such as number of lines of code through time or compilation behavior. But analyzing the structure of the code itself, such as how long conditional statements persist and how often students rename variables, gives a richer description of students' cognitive processes as they write code. Correlating these new descriptive features with academic outcomes allows researchers to empirically determine what best coding practices are, rather than using folk wisdom, and enables measurement of student performance against best practices.

Analyzing the structural evolution of code is difficult. Recent work by ~\citet{brown2024writing} used manual tagging to determine structural statistics such as how often participants comment out code and how often they paste code from outside the environment. While the results are an important step towards a new class of behavioral descriptors, the manual work involved limits scalability.

Obtaining these features broadly and automatically has never been attempted. Some previous work has analyzed character-level changes through time~\cite{edwards2020study,shrestha2022codeprocess}, but this gives very limited information about the code at a structural level. Other work has analyzed evolution of abstract syntax trees~\cite{telea2008code,piech2012modeling}, which has proven to be the most promising approach. However, one key capability is required to unlock full structural analysis: the ability to track a syntax tree node over time. That is, given a node (e.g., an identifier) at time $t$, which node does it correspond to at time $t+i$? Answering this question, without using tree comparison heuristics that are complicated and error-prone, has proven elusive.

In this paper, we present the first algorithm to track parse tree nodes through time. In addition, we present a supplementary algorithm which constructs "bridging" parse trees to represent unparseable code states. We then use these algorithms to partially replicate the work of \citet{brown2024writing} at scale and to perform additional analyses made possible through automated tracking of parse trees. We use a public dataset of keystroke-level logging of CS1 students~\cite{edwards2023review} in our analysis.

\subsection{Contributions and Research Questions}

This paper provides the following contributions:

\begin{itemize}
    \item Development of a novel algorithm for tracking parse tree nodes through keystroke-level code edits, including tracking code that is commented out and then uncommented.
    \item Development of a novel algorithm to create "bridging" parse trees when code is in an unparseable state.
    \item Application of these algorithms to answer basic questions about student programming process.
\end{itemize}
We answer the following research questions using our parse tree node tracking algorithm:
\begin{itemize}
    \item[RQ1] \rqone
    \item[RQ2] \rqtwo
    \item[RQ3] \rqthree
    \item[RQ4] \rqfour
\end{itemize}

\section{Background}

\subsection{Studying student code}

Learning to program involves, among other things, mastering the art of problem solving and the notation and syntax of a programming language~\cite{duBoulay:1986,lahtinen2005study}. Given this, it is understandable that computing education research has extensively examined the outputs of learning programming, such as student-generated code and the errors associated with the code. Student code has been explored from various angles, including looking for misconceptions~\cite{programming-errors}, code quality~\cite{static-analyses-in-py-courses, alamutka-2005-auto-ass-survey}, code complexity~\cite{alamutka-2005-auto-ass-survey}, and specific errors within the code~\cite{becker2019compiler}, among other aspects~\cite{ihantola2015educational}.

Learning to program takes effort~\cite{mccracken2001multi} and learners spend a significant amount of time grappling with basic mechanics~\cite{robins2006problem}. One particularly time-consuming part is programming errors, which takes time from both high-performing and low-performing students~\cite{denny2012all}, from some more than others~\cite{altadmri201537,denny2012all}. There are a wide variety of factors that contribute to learning programming, besides the pedagogy~\cite{vihavainen2014systematic}. As an example, the context in which programming occurs~\cite{becker2019compiler}, the specific programming environment~\cite{denny2011understanding,vihavainen2014novices}, the presence of errors in code templates~\cite{ihantola2015educational,spacco2015analyzing}, the chosen programming language~\cite{stefik2013empirical}, and even the learners' native languages~\cite{reestman2019native} shape what is being learned and how.

\subsection{Analysis of programming process data}

With the increasing availability of automated assessment tools and IDEs that collect data from the programming process, the analysis of programming process data has received more research attention. One milestone in computing education research on this front is the Error Quotient (EQ) metric, which quantified the prevalence of errors in subsequent compilations and consequently students' ability to identify and fix them~\cite{jadud2006methods}. Later work has expanded the idea of looking into subsequent compilations by including the time to fix errors~\cite{watson2013predicting}, considering runtime errors and the use of IDE features~\cite{carter2015normalized}, and looking into the repetition of errors in the programming process in more detail~\cite{becker2016new}. In general, all of these studies highlight that looking into the process of how a program is constructed also reveals information of student performance~\cite{jadud2006methods,watson2013predicting,carter2015normalized,becker2016effective}, again acknowledging the potential context-specificity of these methods~\cite{richards2018investigating,petersen2015exploration,ahadi2015exploring}

While the above examples have primarily focused on analysis of subsequent compilation events or similar events at a similar granularity, recent research has also involved analysis of fine-grained process data such as keystroke logs ~\cite{ihantola2015educational}. Keystroke data can provide insight into how novices learn to construct statements~\cite{vihavainen2014novices}, pauses in programming~\cite{shrestha2022pausing}, how typing of programming constructs evolves over time~\cite{edwards2020programming}, and additional insight into the frequency of compilation errors over the programming process~\cite{vihavainen2014analysis}. 

\citet{brown2024writing} used line edit data (all contiguous edits on the same line are collapsed into one edit event) collected through BlueJ~\cite{bluej} to analyze the frequency of many higher level programming behaviors. They tagged and validated their whole dataset to find how common actions like copy-pasting, renaming variables or commenting/uncommenting code are, as well as how they respond to errors. They discuss in this paper many of the difficulties and limitations of using manual tagging for this task. In our paper, we use parse trees to reproduce many of the same analyses without requiring manual tagging of data, although, as they mention, some of the behaviors, such as the students' error responses, aren't as suitable for automation.

\subsection{Programmatic representation of code}

In programmatic analysis, forming a representation typically starts by splitting the code into tokens and then constructing a parse tree, a hierarchical representation based on the programming language's grammar. Since parse trees can be large, they are often simplified into Abstract Syntax Trees (ASTs), which retain the program's syntactic structure in a more concise, hierarchical form~\cite{aho2007compilers, engineering-compiler}. Depending on the intended use, ASTs can be standardized and pruned. Some systems, like MistakeBrowser~\cite{glassman-reusable-feedback} and Codewebs~\cite{codewebs}, streamline ASTs by removing less important tokens; similarly, the Control-Flow Abstract Syntax Tree (CFAST) is a variant that models a program's control structure by pruning ASTs to only include tokens that are essential for understanding the program's control flow. Other applications involve traversing the AST to form vector representations (e.g. code2vec~\cite{alon2019code2vec}). While most research using tree representations of code has used ASTs, we use parse trees for our analysis. Reasons for this decision are discussed in Section~\ref{sec:ast-vs-pt}.

One alternative to a tree-based representation is \textit{srcML}\cite{collard2013srcml}, an XML representation of source code. \textit{srcML} uses markup tags to designate language constructs instead of nodes. One benefit of this approach is that the XML representation is fully lossless, preserving all source code characters, including whitespace and comments. Parse trees exclude whitespace and comments and abstract syntax trees additionally exclude some syntactic tokens. \textit{srcML}'s approach also allows for the handling of unparseable source code, something we also attempt to solve in this paper. The benefit of our approach is that contextual information about the source code prior to and following the unparseable code state is maintained, enabling well-defined and provable node tracking through time.

\subsection{Research gap}

So far, studies into programming process that use tree-like representations of code have only used summary statistics, such as tree depth or number of nodes of different types. We are aware of no algorithms that track structural changes to the tree. In this paper, we address this research gap, proposing novel algorithms for tracking parse tree changes at every keystroke even through unparseable code states.

\section{Temporal parse tree node hierarchy} 
\label{sec:temporal-hierarchy}

\subsection{Parse trees vs abstract syntax trees}
\label{sec:ast-vs-pt}
We track parse tree nodes through time while most existing research uses ASTs. An AST is a tree representation of code that is compact and easy to understand, but ASTs abstract away many language-specific constructs, such that many tokens in the code are not represented in the tree. Our algorithm depends on correlations between the text representation of the code and the tree, with every character in the code, with the exception of whitespace and comments, being represented somewhere in the tree. Parse trees provide this. See Figure~\ref{fig:temporal-node-hierarchy}.

\subsection{Temporal node hierarchy}
A parse tree has a static hierarchy. For example, an \textit{identifier} node may have a parent of type \textit{operation}, which itself has a parent of type \textit{expression}. This is the hierarchy we normally think of when we think of parse trees. In our work, we use both the static hierarchy and a concept we call the temporal node hierarchy. We define this concept in this section.

Define a "state" as the state of the code after an event that modifies the code, such as a keystroke or paste. Define a "snapshot" as the code at a given state. A temporal parent node $n_{t-i}$ at state $t-i$ is the node from which a node $n_t$ at state $t$ is derived. For example, if an identifier \tt{rati} at state $t-1$ is changed to \tt{ratio} at state $t$, then the node $n_{t-1}$ for identifier \tt{rati} would be the temporal parent of the node $n_{t}$ for \tt{ratio}. The two nodes $n_{t-1}$ and $n_{t}$ do not appear in the same parse tree since they exist at different code snapshots -- they are linked temporally, not statically. See Figure~\ref{fig:temporal-node-hierarchy} for a graphical representation of this temporal parent-child relationship.

Temporal parent nodes are usually identical to their temporal children. This is because from one state to the next, most parse tree nodes don't change. For example, a node for an identifier \tt{sum} won't change from one state to the next if the student is changing some other part of the code. In this case, the temporal parent $n_{t-1}$ and temporal child $n_t$ have identical structures. This doesn't necessarily mean that the subtree under the node doesn't change. For example, a \tt{for} loop node will be identical from one state to the next if a \tt{print} statement in the loop body changes. But, in this case, the subtree under $n_t$ will be different from that of $n_{t-1}$.

While most temporal parent-child node pairs are identical, it is possible for a temporal node parent to be radically different from its child -- even its type may be different. For example, consider a node for a Python identifier \tt{pas}. The node is of type \textit{identifier}. If the programmer then adds an \tt{s}, the lexeme becomes \tt{pass}, which is a keyword in Python that has its own parse tree node type \textit{pass}. So the parent of the \textit{pass} node is a node of type \textit{identifier}.

\subsection{Non-contiguous correspondences}
A temporal node parent may not be in the snapshot immediately preceding the snapshot of its child. This can happen in the case of unparseable states and when commenting out and then uncommenting code. In the case of unparseable states, by definition, no parse tree exists for that snapshot. Consider a node $n_t$ for an identifier \tt{avg}. Suppose that, because of changes unrelated to the identifier \tt{avg}, the snapshots at $t+1$ and $t+2$ are unparseable. But if the snapshot at $t+3$ is once again parseable and the identifier \tt{avg} hasn't changed, then its parse tree node $n_{t+3}$ should have the parent $n_t$, even though there are no parse trees at snapshots $t+1$ and $t+2$, and therefore no corresponding nodes $n_{t+1}$ and $n_{t+2}$. 

The other case of parent-child node pairs that are not temporally contiguous is in the case of commenting out and then uncommenting code. For example, consider the node $n_t$ representing a \tt{print} statement at snapshot $t$. Suppose the \tt{print} statement is commented out at time $t+1$ and then uncommented at the next state at $t+2$. The node $n_t$ is the parent of $n_{t+2}$ even though there is no corresponding node at time $t+1$. Our algorithm tracks this parental relationship as described in Section~\ref{sec:commented-out}.

\section{Parse tree node tracking algorithm}
\label{sec:algorithm}
\begin{figure}
    \centering
    \includegraphics[width=0.5\textwidth]{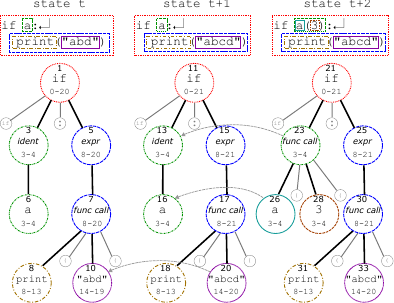}
    \caption{Along the top are snapshots corresponding to three states.
    Below the code snapshots is the parse tree node hierarchy. Each node's outline is the same style as the rectangle outline around the node's corresponding characters in the code snapshot. In each node, the top number is a unique ID and the bottom numbers are the range of indices of the corresponding characters in the code snapshot, where a range of \tt{i-j} indicates the range $[i,j)$. Curved arrows connect nodes with their temporal parents. For clarity, not all nodes and temporal node correspondences are shown.}
    \vspace{-1\baselineskip}
    \label{fig:temporal-node-hierarchy}
\end{figure}




\algnewcommand\algorithmicforeach{\textbf{for each}}
\algdef{S}[FOR]{ForEach}[1]{\algorithmicforeach\ #1\ \algorithmicdo}

\begin{algorithm}
\caption{Parse tree node tracking}
\label{alg:tracking}
\begin{algorithmic}[1]
\Require snapshots $S_t, S_{t-1}, S_{t-2}, ..., S_0$
\Require parse trees $T_t, T_{t-1}, T_{t-2}, ..., T_0$
\Require character correspondences $C_{t-1}, C_{t-2}, ..., C_1$
\LComment{Define $r_t^n$ to be the start and end indices of the range of characters for a node $n$ in snapshot $S_t$. Note that $|r_t^n|=2$.}
\LComment{Define $\hat{r}_{t,u}^n$ to be $[C_u[\cdots[C_{t-1}[C_t[r_t^n[0]]]]\cdots]]$, $[C_u[\cdots[C_{t-1}[C_t[r_t^n[1]]]]\cdots]]$. (Boundary cases described in Section \ref{sec:character-corr}.)}
\If{String s inserted at index $i$}
    \State $C_t \gets [0,1,\cdots,i-1] + [-1]^k + [i,i+1,\cdots,|S_{t-1}|]$ where $[-1]^k$ is an array of $-1$s of size $|s|$.
\ElsIf{String s deleted at index $i$}
    \State $C_t \gets [0,1,\cdots,i-1,i+|s|,i+|s|+1,\cdots,|S_{t-1}|]$
\EndIf
\ForEach{node $n \in T_t$}
    \State $i \gets 0$
    \Repeat
        \State $i \gets i+1$
        \State Find node $m \in T_{t-i}$ s.t. $\hat{r}_{t,t-i}^n \subseteq r_{t-i}^m$ and other conditions described in Section \ref{sec:simple-node-correspondence}
    \Until{$m$ exists or $\hat{r}_{t,t-i}^n = (-1,-1)$}
\EndFor

\end{algorithmic}
\end{algorithm}

The temporal node hierarchy for parse trees is computed using Algorithm~\ref{alg:tracking}, which proceeds in two steps. At each state, the first step is to compute character correspondences (lines 3-6). This is a mapping of characters in a snapshot to where the characters were in a previous snapshot. The second step is to use those character correspondences to find the node correspondences (lines 7-12). The $+$ operator on line 4 is array concatenation and the subset notation on line 11 is defined as $m[0] \ge n[0]$ and $m[1] \le n[1]$ for ranges $m$ and $n$.

In Fig.~\ref{fig:temporal-node-hierarchy} we see the snapshots for three states. We use these states for a running example demonstrating the algorithm.

\subsection{Character correspondence}
\label{sec:character-corr}
\begin{figure}
    \centering
    \includegraphics[width=0.45\textwidth]{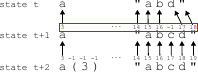}
    \caption{Character correspondence arrays for the series of states in Figure~\ref{fig:temporal-node-hierarchy}.}
    \label{fig:char-correspondence}
    \vspace{-1\baselineskip}
\end{figure}

We maintain a character correspondence array $C$. For each snapshot of length $n$ characters, there is one $C_t$ array such that if a character in position $i$ was at position $j$ at the previous snapshot $t-1$, then $C_t[i]=j$. See Figure~\ref{fig:char-correspondence}, where a subset of the $C_{t+1}$ array for snapshot $t+1$ is highlighted in the box.

The array is computed as shown in lines 3-6 of the algorithm. In keystroke data, each event is either an insertion of contiguous characters or a deletion of contiguous characters. For an inserted character at index i (line 4), its correspondence to the previous snapshot is $C_t[i]=-1$ since there is no corresponding character in the previous snapshot. For deleted characters, they no longer exist in snapshot $t$, and so they have no entry in $C_t$.

We can track a character correspondence back as many states as needed. Line 2 of the algorithm shows a chained representation of this, where we find $\hat{r}_{t,u}^n$, which is a pair of numbers representing the character range of a parse tree node $n_t$ as it appeared in the code at state $u$. That is, if the characters of an identifier $foo$ at state $u$ are at indices $r_u^n=[12,15)$ but then at state $t$ a character is inserted at the very beginning of the file, then the new index range at state $t$ is $r_t^n=[13,16)$. Using the character correspondence array, we can track from state $t$ back in time to state $u$ to find $\hat{r}_{t,u}^n=[12,15)$. Boundary conditions exist when either the start or end index in the range corresponds to $-1$. In this case, we search forward or backward in the array from the start or end index, respectively, to find a value that is not $-1$. In the case that both the start and end indices correspond to $-1$ then then entire string is new (such as in the case of a paste) and there is no correspondence.

The character correspondence array $C$ allows us to answer the question, "where was this character in a previous snapshot?" Previous approaches to answering this question used diffs \cite{fujimoto2021towards}, which are heuristic, but we have the advantage of knowing exactly what changes are happening with keystroke data.

\subsection{Simple node correspondence}
\label{sec:simple-node-correspondence}
Fig. \ref{fig:temporal-node-hierarchy} shows how we determine the temporal node hierarchy where $n_{id}$ indicates the node with ID $id$. Consider node $n_{20}$ ("abcd" in snapshot $t+1$). The parse tree computation code (our implementation uses the ANTLR library~\cite{antlr}) gives us the range of characters $r_{t+1}^{n_{20}}=[14,20)$. Using the character correspondence array (Fig.~\ref{fig:char-correspondence}) we can determine that that range of characters came from the range $\hat{r}_{t+1,t}^{n_{20}}=[14,19)$ in snapshot $t$. We then search the parse tree at snapshot $t$ for the node $n_j$ with the smallest range $r_t^j$ under the constraint that $\hat{r}_{t+1,t}^{n_{20}} \subseteq r_t^j$. To find $j$, we first test against the root node $n_1$ which passes since $\hat{r}_{t+1,t}^{n_{20}}=[14,19) \subseteq [0,20)=r_t^{n_1}$. We search deeper in the tree looking for a smaller node and test $n_5$, then $n_7$, and finally $n_{10}$: $\hat{r}_{t+1,t}^{n_{20}}=[14,19) \subseteq [14,19)=r_t^{n_{10}}$. $n_{10}$ is a leaf node, so we set the the parent of $n_{20}$ to be $n_{10}$. We do not have to test every node in the parse tree at snapshot $t$: if we test some node $n_j$ and the ranges are disjoint then we do not have to test descendants of $n_j$.

Now consider node $n_{23}$ in snapshot $t+2$. Using the same algorithm we find that it corresponds to node $n_{13}$ in snapshot $t+1$. The power of this algorithm is demonstrated by the fact that these two nodes correspond, yet they are of different types: $n_{13}$ is a variable identifier and $n_{23}$ is a function call, yet with this correspondence algorithm we know that $n_{23}$ was derived, or created from, $n_{13}$, something we would not have discovered had we just been doing tree comparisons.

\subsection{Commented-out code}
\label{sec:commented-out}
Suppose at state $t+1$ we comment out a \tt{print} statement, then restore it at state $t+j$. The \tt{print} call does not appear in the parse tree at states $t+1$ through $t+j-1$, so it may appear that the node at state $t+j$ has no parent. However, the character correspondence array indicates that the characters do exist in the previous snapshot, suggesting that the code is now in a comment, thus not appearing in the parse tree. Line 11 of Algorithm~\ref{alg:tracking} is where a previous parse tree is searched for a node corresponding to a node of interest in the current snapshot. In most cases this will occur in the previous snapshot. But in cases where the code was in a comment in the previous snapshot $t+j-1$, the characters in snapshot $t+j-1$ that correspond to the node in snapshot $t+j$ are in a comment, and thus are not represented in a leaf node of the parse tree $T_{t+j-1}$. We iterate until we find a snapshot where the characters are represented in the parse tree or until the characters disappear (in which case the correspondence range is (-1,-1)).


\section{Bridging parse tree construction algorithm}
\label{sec:bridging}

In this section we describe the process for creating bridging parse trees for unparseable code states using Algorithm \ref{alg:bridging}. The idea is to create, for an unparseable snapshot, a tree that is as much like a parse tree as possible. We call these bridging parse trees (BPTs) because they provide a bridge across unparseable snapshots. The structure of bridging parse trees does not conform to the grammar of the language, but all node types come directly from the grammar, with the exception of some nodes we call transient nodes, discussed in Section~\ref{sec:transient}. We construct BPTs using a combination of the trees of both the previous state and the next parseable state. This algorithm must account for three types of code edits: insert edits that exist in the next parseable state (lines 5-15), insert edits that are deleted before the next parseable state, which we will call transient edits (lines 16-19), and delete edits (lines 20-21). We assume that the parse tree node tracking algorithm described earlier has already been applied to all parseable snapshots.

\begin{algorithm}
\caption{Construct bridging parse trees for all unparseable states}
\label{alg:bridging}
\begin{algorithmic}[1]
\Require snapshots $S_{0}, S_{1}, ... S_{n-2}, S_{n-1}$
\Require parse trees $T_t$ for all $t$ s.t. $S_t$ is parseable 
\Require character correspondences $C_{0}, C_{1}, ... C_{n-2}, C_{n-1}$
\LComment{Define $r_t^n$ to be the start and end indices of the range of characters for a node $n$ in snapshot $S_t$.}

\ForAll{$t \gets 1$ to $n-2$}
    \If{$S_t$ is unparseable}
        \State{$T_t \gets$ copy($T_{t-1}$)}
        \If{String $c$ is inserted at index $i$ of $S_t$ and $c$ corresponds to $S_{t+m}$ where $S_{t+m}$ is the next parseable snapshot}
            \State $i_{t+m} \gets $ the location of $i$ in state $t+m$
            \State Traverse $T_{t+m}$ searching for the lowest node $s$ s.t. $i_{t+m} \subseteq r_s$
            \State $s' \gets$ a copy of the highest ancestor of $s$ with no temporal parent or a copy of $s$ if none exists
            \State Prune $s'$
            \ForAll{$i_k\in r_{t+m}^{s'}$ where $C_{t-1}[C_{t}[\cdots[C_{t+m-1}[C_{t+m}[i_k]]]] \neq -1$}
                \State Prune $T_t$ of the character at $i_k$
            \EndFor
            \If{$s'$ does not have a temporal parent}
                \State Insert $s'$ as a child of the temporal parent of the parent of $s'$
            \Else
                \State Replace the temporal parent of $s'$ with $s'$
            \EndIf
        \ElsIf{String $c$ is inserted at index $i$ of $S_t$ and $c$ does not correspond to $S_{t+m}$}
            \State $s \gets$ a new transient node, with range $r=[i,i+|c|+1)$
            \State $p \gets$ the highest, leftmost node in $T_t$ s.t. $r_p[0]\ge r_s[0]$ 
            \State Insert $s$ as an older sibling to $p$
        \ElsIf{String $c$ is deleted at index $i$ of $S_t$}
            \State Prune $T_t$
\EndIf\EndIf\EndFor
\end{algorithmic}
\end{algorithm}

When using trees $T_{t-1}$ and $T_{t+m}$ to construct a bridging parse tree $T_t$ for state $t$, we use the character correspondence array $C$, described in Section \ref{sec:character-corr}, to determine what pieces of $T_{t-1}$ and $T_{t+m}$ exist in snapshot $S_t$. Say for snapshot $S_{t-1}$, we have some character $c$ at index $i$. If $i\in C_t$ (some character in state $t$ corresponds to $c$ in state $t-1$) then we say that $c$ corresponds to $S_{t-1}$.

\subsection{Pruning subtrees}
\label{subsec:pruning}
Both insert and delete edits rely on pruning characters from a subtree (lines 9, 11 and 21). When constructing a bridging parse tree $T_t$ for state $t$, some subtree $s'$ may be inserted into $T_t$. This subtree $s'$ may contain characters that are not present in snapshot $S_t$. 

For example, let $s'$ be a copy of $s\in T_{t-1}$. Subtree $s'$ contains the consecutive set of characters $D_{t-1}=\{c_i, c_{i+1}, ..., c_{i+n}\}$, where $c_i$ and $c_{i+n}$ are the first and last characters in $s'$ as they appear in $T_{t-1}$. For each character $c_i\in D_{t-1}$ we check whether it corresponds to a character in state $t$. If no correspondence exists (i.e., if $c_i$ does not appear in snapshot $S_t$), then that character must be removed from $s'$. For each node $n'\in s'$, with character range $r$, if $i\in r$, then the length of $n'$ is reduced by decreasing the upper bound of $r$. Additionally, if $i<r_l$, where $r_l$ is the lower bound of $r$, both bounds of $n'$ are decreased by one to adjust for the changes in preceding nodes. For example, suppose subtree $s$ corresponds to code at $[10,16)$ in snapshot $S_{t-1}$ and the character at index $13$ does not exist in snapshot $S_t$. In that case, we update $s'$ to have the range $[10,15)$. Descendant nodes of $s'$ are updated similarly.

\subsection{Insert edits}
This and the following two subsections give a narrative explanation of Algorithm~\ref{alg:bridging}.
For some insert edit of string $c$ in the code snapshot $S_t$, we use the parse tree or bridging parse tree $T_{t-1}$ as a starting point, as it incorporates every character in snapshot $S_t$ except for those found in $c$. We assign tree $T_t$  to be a copy of $T_{t-1}$. Using the character correspondence array, we can find the position of $c$ in the next parseable snapshot $S_{t+m}$. We refer to this position as $i_{t+m}$. We then traverse the parse tree $T_{t+m}$, searching for $i_{t+m}$ until reaching a node $s'$ with no temporal parent (i.e. a node created after snapshot $S_{t-1}$). $s'$ is then inserted into $T_t$ as a child of the temporal parent of the parent of $s'$.

However, since there could be multiple unparseable states before state $t+n$, there could be nodes or characters in subtree $s'$ that were inserted after state $t$. To solve this, we prune $s'$ to only include characters with correspondence to state $t$ as described in section \ref{subsec:pruning} (line 9). Another issue is that an insert could affect the structure of existing nodes in tree $T_{t-1}$. We know that any changes caused by the insert $c_t$ would be included in the subtree $s'$, so as shown in lines 10-11, $s'$ is searched for any characters that correspond to snapshot $S_{t-1}$ (i.e. characters that existed in snapshot $S_{t-1}$ but were structurally moved in snapshot $S_t$). See Figure \ref{fig:BridgingTree} for an example of this occurring. These characters are pruned out of $T_t$ (which was copied from $T_{t-1}$), as they will be reinserted inside $s'$.

Alternatively, if the traversal ends, finding the lowest node $s'$ containing $c$, and $s'$ does have a temporal parent, then the edit $c$ occurred inside an existing node in $T_{t-1}$. The same pruning of $s'$ and $T_{t-1}$ occur, and the temporal parent of $s'$ is replaced by the pruned subtree $s'$ (line 15).

\subsection{Transient edits}
\label{sec:transient}
Some characters are inserted and then deleted in a stretch of unparseable states, meaning those characters never appear in a proper parse tree. We refer to these as transient edits. Since transient edits never occur in a parseable state, no information is available as to the edits' placement in the tree, or the type of the node containing the edit. To represent these edits, we create a transient node, which is added into $T_t$, a copy of $T_{t-1}$. The transient node is inserted as a child of the highest level node $p$ whose character range $r$ includes the indices of $c$ (lines 18-19). Inserting the transient node under $p$ is a heuristic choice, as a more precise placement requires unavailable information.

\subsection{Delete edits}
Insert edits are able to use $T_{t-1}$ as a good basis for $T_t$, as $T_{t-1}$ represents every character in $S_t$ with the exception of the inserted character. For deletion edits, ideally the next parseable tree $T_{t+m}$ would be used as a basis for $T_t$, as it would include the most context for what happens to the code without the deleted character. However, since there could be any number of unparseable states before snapshot $S_{t+m}$ that could drastically change $T_{t+m}$, $T_{t+m}$ cannot be used as a basis for deletion in the same way $T_{t-1}$ is used for insertion. We could generate bridging parse trees in reverse, starting from the final state and working backward, using $T_{t+1}$ as the basis for constructing $T_t$. However, this would cause insertions to face the same problems that deletions have. Since insertions account for a higher percentage of edits in our dataset, we chose to prioritize them.

Instead, for the deletion edit of string $c$ in snapshot $S_t$, we let $T_t$ be a copy of $T_{t-1}$, and prune $T_t$ as shown in section \ref{subsec:pruning}, essentially creating a copy of $T_{t-1}$ with $c$ removed.

\subsection{Limitations with bridging}
\label{BridgeIssues}
\journal{During insertion edits, when inserting a pruned subtree from the next parseable state, some nodes can be inserted into the tree earlier than they would exist were it parseable. For example, say in a series of unparseable states, a student writes a print statement, and then writes an if statement around the print statement. In the next parseable tree $T_{t+n}$, nodes for the print statement are descendants of an if statement node. When searching for the subtree to insert the print statement, the first node that has no temporal parent is actually the if statement node. This node is $s$ and will be pruned inserted into $T_t$. $s$ is an if statement node whose only child is the body node containing the print statement. Thus the if statement node is inserted well before the if statement was written. While this does represent the context the print statement will eventually exist in, it doesn't accurately represent the snapshot $S_t$.}


Transient edits are a limitation when used for analysis since they have no anchor in compiled parse trees. $4.2$\% of all edits have one or more transient characters, most of which are deleted shortly after they are written.

\journal{Under unusual conditions, our bridging algorithm constructs counterintuitive trees. One example is when, in a series of unparseable states, students write code in nested scopes.}
In rare cases, our algorithm fails to construct a bridging parse tree. There are a number of diverse causes for these failures, and improving the success rate of the bridging algorithm will be a matter of identifying and correcting these edge cases. Our implementation addresses the most common of these edge cases. In our test dataset, 99.5\% of attempts successfully construct bridging parse trees. If our algorithm fails to construct a bridging parse tree for an unparseable snapshot, BPTs for all following unparseable snapshots cannot be constructed until a parseable state is encountered. Because of these chains of unattempted BPT constructions, $63.0\%$ of all unparseable snapshots result in a BPT. This percentage varies heavily between files (mean$=74.3\%$,\ $\sigma=28.9$).

\section{Experiment methodology}
\subsection{Data} \label{sec:Data}
Data for this research was gathered by tracking the keystrokes from undergraduate students at a university in the United States and were released as a publicly available dataset \cite{DVN/BVOF7S_2022, edwards2023review}. As students worked on their programming homework assignment in Python, the \textit{PyPhanon} plugin to the \textit{PyCharm} Integrated Development Environment captured the changes they made to their code files. These edits were then converted to the \textit{ProgSnap2} data format and used for analysis. Since each file began from an empty state,  the edits can fully recreate the history of the assignment, providing a snapshot of the code at each event.

Over 1 million keystrokes are included in the dataset, collected from 44 students in a CS1 course in the fall of 2021. Data was collected for eight weekly programming assignments, each with 1-3 tasks. The tasks were a mix of turtle graphics drawings and text-based logic problems. There was, on average, $767$ events per student file. 

We filtered out files with fewer than 300 events, as these files were very short, and typically were either a student experimenting briefly or pasting in a final solution that was written in another text editor. After applying our algorithms to derive bridging parse trees for unparseable states, we also discarded $87$ files that had trees (whether parse trees or BPTs) for less than $80\%$ of code states. See Section \ref{BridgeIssues}. To ensure that our subset of retained files did not misrepresent the whole dataset, we ran t-tests on outcomes between the excluded and retained files. There was no detectable difference in assignment score ($t=-0.12$, $p=0.90$), final exams ($t=0.50$, $p=0.61$) or final grade ($t=0.21$, $p=0.83$).  When considering students that had a higher file exclusion rate ($>33\%$), there was similarly no significant difference in assignment score ($t=-0.70$, $p=0.48$), final exam ($t=0.16$, $p=0.88$) or final grade ($t=0.04$, $p=0.96$). There was also no correlation between the number of keystrokes in a file and its percentage of trees ($r=-0.08$, $p=0.11$). Our final dataset was composed of $274$ files from $42$ students.

\subsection{Measures}
\subsubsection{Node deletion rate:} \ 
Define node deletion rate as the percentage of parse tree nodes that are eventually deleted: $\frac{numDeleted}{numNodes}$ where $numDeleted$ is the number of deleted nodes and $numNodes$ is the number of unique nodes. For example, if the starter code was a print statement with a single string as an argument and the only thing the student did was delete the string, then the node deletion rate for nodes under the print statement would be $100\%$, while the rate for the entire program would be $33\%$, assuming there was a single root note as parent of the print node.

\subsubsection{Context switch frequency:}\  \label{JumpingMetric}
In order to answer RQ4, we use a measure that we call context switch frequency to describe how often students jump around in their code. In a parse tree, we use the length of the path between the previously edited leaf node and the current edited leaf node to measure the distance jumped. Whitespace and comment edits are not represented in parse tree leaf nodes and are skipped for these calculations. 

We suggest that using a parse tree to measure jump distance is more meaningful than using the location of text edits in the code. For example, a student may work on two areas that are very close in the source code but very far apart intuitively, as represented by the distance between parse tree nodes. For example, code near the end of one function might be only a few characters away from code in a completely unrelated function. Being close in the source code might indicate that the student is working on related code, but the node distance in the parse tree will correctly suggest that the student has switched contexts.

Using this measurement of jump distance, let $S$ be the set of all jump distances between consecutive edits for some file. Let $S_1=\{x\in S\ |\ x\ge1\}$ and $S_5=\{x\in S\ |\ x\ge5\}$. We define context switch frequency as $\frac{|S_5|}{|S_1|}$. We use $5$ as a jump distance that is big enough to be considered non-sequential coding, even though in some cases, a path length of $5$ edges could still be in the same line of code. So, of the edits that move to a different node, context switch frequency is the percentage that jump a distance of $5$ or more.

\subsection{Comparative results}
We use a number of measures of student behavior recently proposed by \citet{brown2024writing} and compare our results to theirs. This portion of our work serves to generalize their results in some cases, and show differences due to context in others. There are three main differences between our methodology and theirs. First, both datasets were of novice programmers, but ours was a single group of students in the same class, given the same instructions and the same requirements, while theirs incorporated the data of many users of the web IDE BlueJ, incorporating both programming for various course assignments and personal use. Second, we did not rely on manual coding of snapshot evolution. And third, their events were line-level while ours are keystroke-level. Table \ref{ComparisonTable} shows a comparison of our findings with those of \citet{brown2024writing}. There are some differences in our definitions of these behaviors and theirs. Table \ref{fig:definitions} shows the definitions we use. Noted are ways our definitions vary from the prior work.

\begin{figure}
    \begin{minipage}{0.45\textwidth}
        \begin{tabular}{p{7cm}}
            \hline
            \multicolumn{1}{c}{\textbf{Code construction}} \\
            \hangindent=2em Duplication: The user copies existing code they have already written within the project and pastes it elsewhere, making necessary changes to adapt to the new context. \\
            \hangindent=2em Pasting: The user pastes in any amount of code that doesn't appear in any previous snapshot.$^1$ \\
            \hangindent=2em Commenting: The user writes a comment in their code.$^2$ \\
            \hangindent=2em Deleting a comment: The user fully deletes a comment in their code that they had previously written.$^4$ \\
            \hline
            \multicolumn{1}{c}{\textbf{Code editing}} \\
            \hangindent=2em Commenting Out: The user comments out any amount of code.$^1$ \\
            \hangindent=2em Moving: The user moves existing code they have written into another place within the program or project. \\
            \hangindent=2em Renaming: The user changes the names of variables, parameters or methods. At least twenty consecutive edits have to be done on other nodes before returning to edit an \textit{identifier} node to be counted as renaming.$^3$ \\
            \hline
        \end{tabular}
        \captionof{table}{Final definitions of the behaviors used in section \ref{RQ2}. Any variations between our definitions and the tags defined by \citet{brown2024writing} are noted. \\
        $^1$ Brown et al.'s definitions considered edits of a line or more of code, while ours also includes edits of less than a line. We use these definitions as we have access to keystroke-level data, while theirs used line-level data. \\
        $^2$ We combine post- and pre-commenting into one commenting definition, as distinguishing between them would require manual tagging to determine what code a comment is in relation to. \\
        $^3$ We count each instance of the same identifier being renamed, where Brown et al. counted per unique identifier. This will lead to higher number of occurrences, but not effect the occurrence frequency. Brown et al. also excluded variables that were declared but not utilized, while we include these. \\
        $^4$ Deleting comments was not a behavior measured by Brown et al}
        \label{fig:definitions}
        \vspace{-2\baselineskip}
    \end{minipage}
\end{figure}

\begin{figure}
    \centering
    \includegraphics[width=0.45\textwidth]{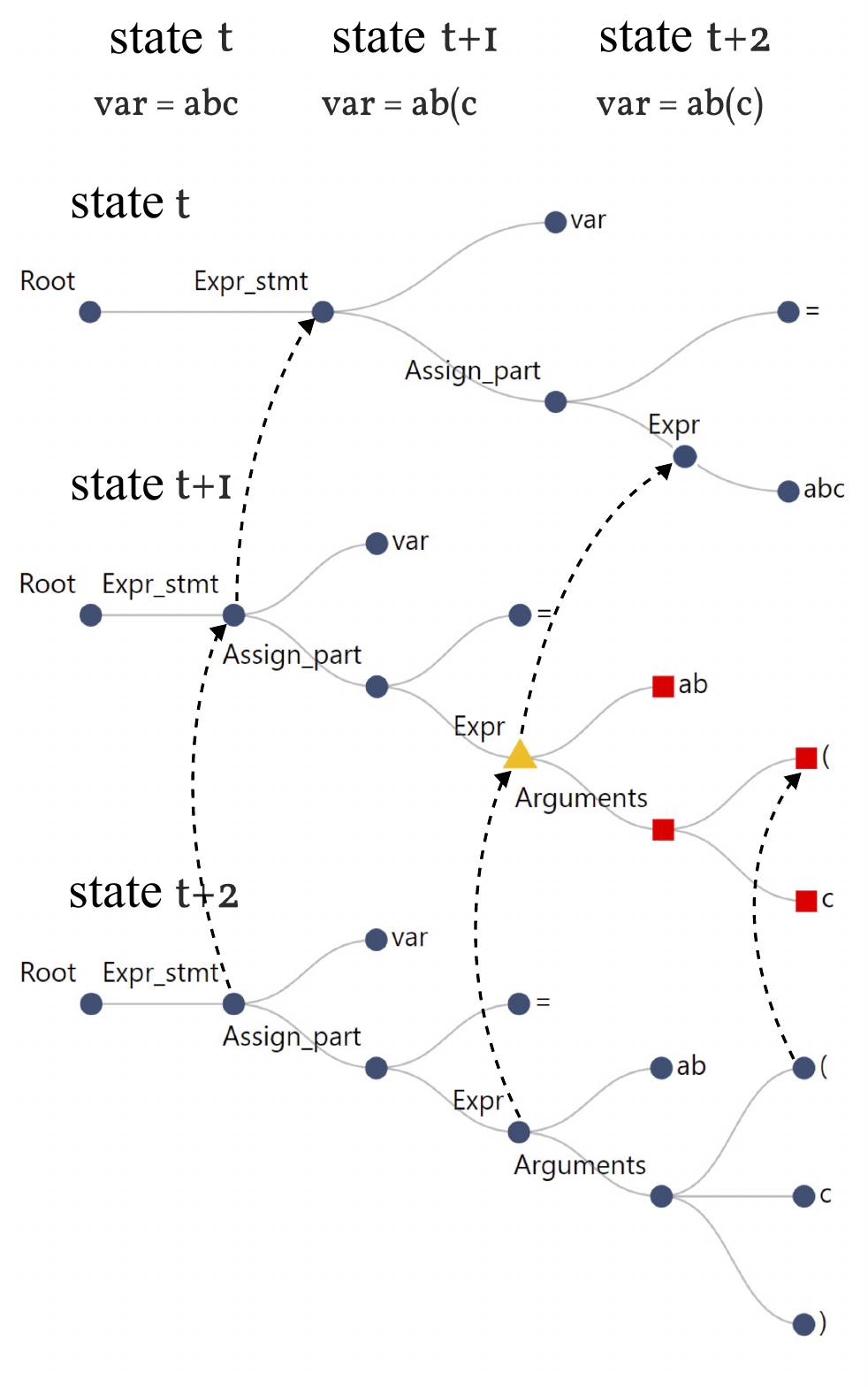}
    \caption{Node tracking example. Three trees are shown for three consecutive code snapshots. State $t+1$ is unparseable, so the bridging tree at state $t+1$ was derived from state $t$ and $t+2$. Red square nodes were inserted or changed from the tree $t$. The yellow triangle node has a temporal parent in the tree $t$, but had its list of children changed. Dotted lines indicate some of the temporal relations between nodes in these states.}
    \label{fig:BridgingTree}
    \vspace{-1\baselineskip}
\end{figure}

\section{Results and discussion}
\subsection{RQ1}
\subsubsection{Node tracking:}
\begin{figure}
    \centering
    \includegraphics[width=0.45\textwidth]{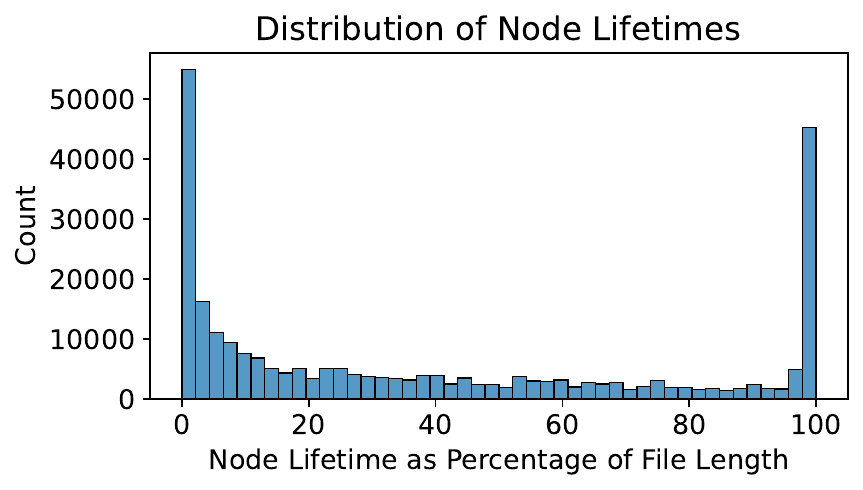}
    \caption{Distribution of all node lifetimes. Node lifetime is the percentage of snapshots in a file a node exists for.}
    \label{fig:lifetimes}
\end{figure}

\ 
Our first research question asks, "\rqone" Figure~\ref{fig:BridgingTree} shows an example of our algorithm in practice. Note that node correspondences are maintained across multiple states and can continue through unparseable states. The algorithm scales to submissions with many states and large parse trees. The largest parse tree in our dataset has $4473$ nodes and the deepest tree has a depth of $60$ nodes. The longest persisting node in our dataset exists through 7388 states and 20\% of all nodes persist for at least 1092 states. Figure~\ref{fig:lifetimes} shows a distribution of the percentage of time that nodes exist for. The spike in number of nodes lasting nearly $100\%$ of the time are due to starter code that is pasted in early and never deleted. The largest spike around $0\%$ is due to nodes that are deleted shortly after being written. A uniform distribution of the remainder would have indicated that students consistently progress toward the solution, but because nodes are deleted during development, we gradually see fewer nodes showing longer persistence.


\journal{
\begin{figure}
    \centering
    \includegraphics[width=0.49\textwidth]{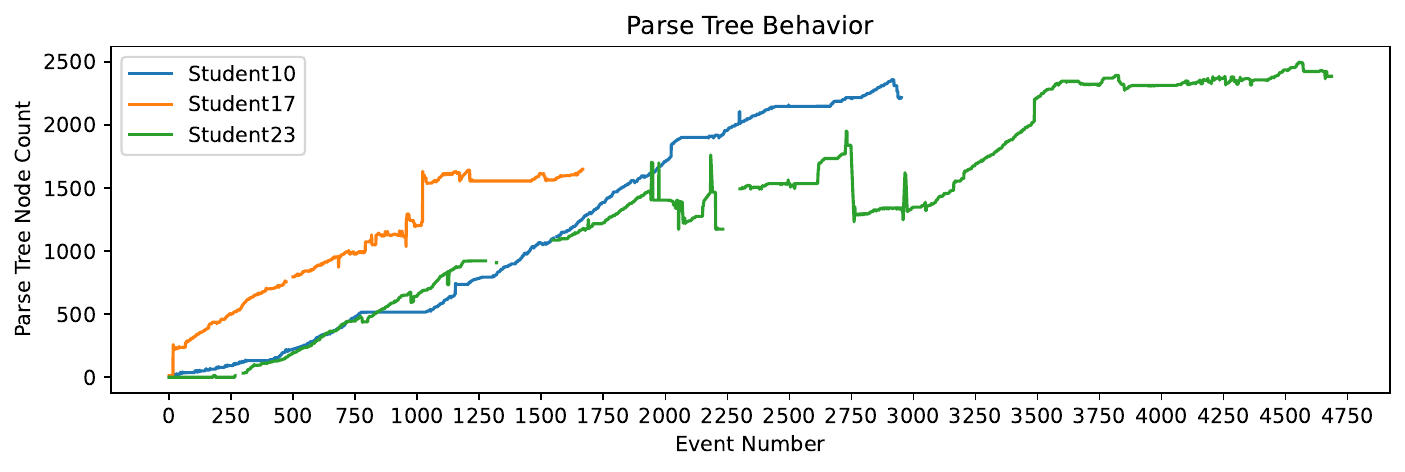}
    \caption{Evolution of parse trees for three students on the same assignment. Each "line" shows the size of the parse tree at each keystroke. Gaps indicate where BPT construction failed. The thing to note here is the very different behaviors of the parse trees across students.}
    \label{fig:progress}
\end{figure}
}

\journal{
Figure~\ref{fig:progress} shows the progress three different students made on a programming assignment. The difference in progress toward completion is evident. Some students make steady progress whereas other students' behavior is less consistent. This chart includes states that are unparseable. In those cases, the number of nodes at a given state is given by the number of nodes in the bridging parse tree.
}

$6.3$\% of nodes are commented out at some point and $30.9$\% of those are eventually restored. $39.8$\% of all nodes are eventually deleted, but the behavior varies widely across students and assignments. Grouped by assignment, $21.5$\% are deleted in assignment 11 and $56.8$\% are deleted in assignment 6. Student behavior differs widely as well: the average student deletes $37.4$\% of all nodes, but the standard deviation is $19.4$\%.

\subsubsection{Bridging parse trees:} \ 
Figure \ref{fig:BridgingTree} presents an example of our bridging algorithm. In state $t+1$, the algorithm copies $T_t$ as $T_{t+1}$.
\journal{It then searches tree $T_{t+2}$ for the inserted character "(", and finds the "\textit{Arguments}" node, which is the first node that does not have a temporal relation to tree $T_t$. It prunes the \textit{Arguments} node to only characters in state $t+1$, removing the ")" child. In state $t+2$, \textit{Arguments}' parent is the \textit{Expr} node, which has a temporal parent in state $t$. The pruned \textit{Arguments} node is then inserted as a child of this \textit{Expr} node in $T_{t+1}$. Lastly, the inserted nodes are searched for any characters that correspond to state $t$, and finds that the character in the "\textit{c}" node was a part of the node "\textit{abc}" in $T_t$. To account for this, the size of the \textit{abc} node is decreased accordingly.}
It then searches tree $T_{t+2}$ for the inserted character "(" and adds a pruned version of the supporting "\textit{Arguments}" subtree.
This shows the algorithms' ability to create BPTs, even when nodes were split or changed ("abc" to "ab" and "c"). 


As discussed in Section \ref{BridgeIssues}, the majority of unparseable snapshots lacking a bridging parse tree are due to chained failures of contiguous unparseable states. Given a valid tree $T_{t-1}$, the following tree $T_{t}$ is correctly created $99.5\%$ of the time. In our dataset, an average file is parseable $56\%$ of time. Constructed BPT's added trees for an additional $29\%$ of code states on average. $274$ of the $369$ student files we tested had a parse tree or BPT for at least $80\%$ of states.



\subsection{RQ2} \label{RQ2}
Our second research question asks, "\rqtwo" To answer this question, we use the combination of a keystroke event dataset and the sequences of parse trees we've created to programmatically recognize and track the occurrences of many of the behaviors that were studied by Brown et al. \cite{brown2024writing}. We report the frequencies of behaviors, such as copy-pasting, writing comments, commenting/uncommenting code, and renaming variables. We also look at deletion in more detail, which will be discussed in section \ref{RQ3}. Brown et al. divided their files into short (300-500 events) and long (500-1000 events) files, while excluding files of more than 1000 events because of the labor of manual tagging. Their dataset is collected at line-level edits, while ours is keystroke level. Therefore, one event is not the same in both contexts. In our dataset, on average, a student makes 9.94 edits per line. While there is a large variance ($\sigma=17.2$), this will be our comparison for file sizes. For example, a short file would be ~3000-5000 keystroke events. Most assignments in our dataset consisted of multiple shorter tasks, so our dataset has only 3 files that are in Brown's long category. For this reason, we will relabel Brown's short files as medium, and will describe our findings using the notation  \textit{Short | Medium} for files with <3000 | 3000+  respectively. In our dataset, there were $233$\ |\ $41$ files of each size. \citet{brown2024writing} analyzed approximately 56,000 line level events while we analyzed 466,000 keystroke events.

\begin{table}
    \centering
    \begin{center}
        \begin{tabular}{|w{l}{3cm}|c|c|c|c|}
        \hline
        \multicolumn{1}{|c|}{Events} & \multicolumn{2}{|c|}{Brown} & \multicolumn{2}{|c|}{Ours} \\
        \hline
        & Md \% & Lg \% & Sh \% & Md \% \\
        \hline
        Exterior paste & 15 & 19 & 70 & 83 \\
        \hline
        Code duplication & 54 & 78 & 58 & 88 \\
        \hline
        Moving code & 25 & 39 & 15 & 34 \\
        \hline
        Renaming variables & 12 & 14 & 37 & 71 \\
        \hline
        Commenting code & 6 & 6 & 53 & 80 \\
        \hline
        Uncommenting code & 8 & 8 & 42 & 76 \\
        \hline
        Writing comments & 18 & 28 & 50 & 88 \\
        \hline
        Deleting comments & & & 43 & 78 \\
        \hline
        \end{tabular}
    \end{center}
    \caption{Comparison between Brown et al's \cite{brown2024writing} findings, and the finding presented in this paper, as to the frequencies of programming behaviors. All numbers are the percentage of files with at least one instance of the corresponding behavior.} 
    \vspace{-2\baselineskip}
    \label{ComparisonTable}
\end{table}

Pasting from an exterior source was fairly common, as $77\ |\ 88\%$ of files had at least one occurrence. However, many of the assignments including provided starter code that was pasted in by the student. We control for this by excluding any paste that is a significant substring of any starter code. Removing starter code pastes reduces the frequency of exterior pastes slightly to $70\ |\ 83\%$ over all files, with median occurrences of $2\ |\ 4$. Code duplication, or copy-pasting within the same file was slightly less common, in $58\ |\  88\%$ of files, with median occurrences $2\ |\ 4$ per file. Moving code, or cut-and-pasting, was more rare ($15\ |\  34\%$, median occurrences $1\ |\ 1.5$).

Renaming variables was fairly common ($37\ |\  71\%$, median occurrences $2\ |\ 4$). Commenting out and uncommenting code were both reasonably common ($53\ |\  80\%$ and $42\ |\ 76\%$  respectively). $22\%$ of instances of commenting out code included commenting out a print statement. We suggest that most of these print statements were used for debugging. Writing comments was also fairly common, in $50\ |\  88\%$ of files, median occurrences of $5\ |\  9.5$. We're unable to make a distinction between pre- and post-commenting automatically. Deleting comments was slightly less common, with $43\ |\  78\%$, but had lower median occurrences of $2\ |\ 4$.

Many of these statistics don't line up with Brown's findings. This is likely due to contextual differences of the datasets. For example, we found much higher rate of pasting from external sources.  One contextual difference in datasets is that not all BlueJ users are writing code for an assignment where they must meet some criteria within a time frame. Some are just experimenting, or recreationally programming. This could lead to lower occurrence rates of behaviors like commenting and uncommenting code, renaming variables, and moving code, which we suspect would be more common during the debugging process. We also found a much higher rate of writing comments, possibly because writing comments was an expectation or requirement in the course, whereas recreational codes might be expected to have fewer comments. Differences in the datasets' level of granularity also accounts for some of the increase in behavior frequencies, as we can detect behaviors that occur during the writing of a line. For example, if a student copy-pastes a section of a line of code, our dataset shows that as one 'event', while in the BlueJ line-level dataset, a partial line paste is undetectable as part of a larger line edit. Other contextual differences between the datasets such as the programming language used (Python vs Java) could also account for some difference in findings. 

There are a number of Brown et al.'s tests that we couldn't replicate. Primarily, looking at student's error responses isn't practical. While compiler errors could possibly be used to make some inference into the context behind errors, there is a lot more nuance to the labeling of error response that requires manual tagging. Brown also studies larger scale coding strategies, for example "sequential" coding, (writing code in order, top to bottom). We discuss our  approximate measurements of sequential coding in Section~\ref{RQ4}. Despite the limitations, there are also many interesting benefits to using trees, including using node count as a measurement of the size of an edit.

One final analysis is that of progress in terms of parse tree size (Figure~\ref{fig:progress}). Analyses have been done using the development of Abstract Syntax Trees \cite{feist2016visualizing}, but none at the keystroke granularity. Additionally, our analysis has roughly 30\% more data because we can use bridging parse trees to fill in gaps at unparseable code states. The thing to note in this chart is the large differences in how students progress toward a solution. Student10 shows steady progress while Student23 deletes or comments out large swaths of code around events 200, 2100, and 2750. The deleted code after 2750 is slowly rebuilt over 500 events.

\begin{figure}
    \centering
    \includegraphics[width=0.8\columnwidth]{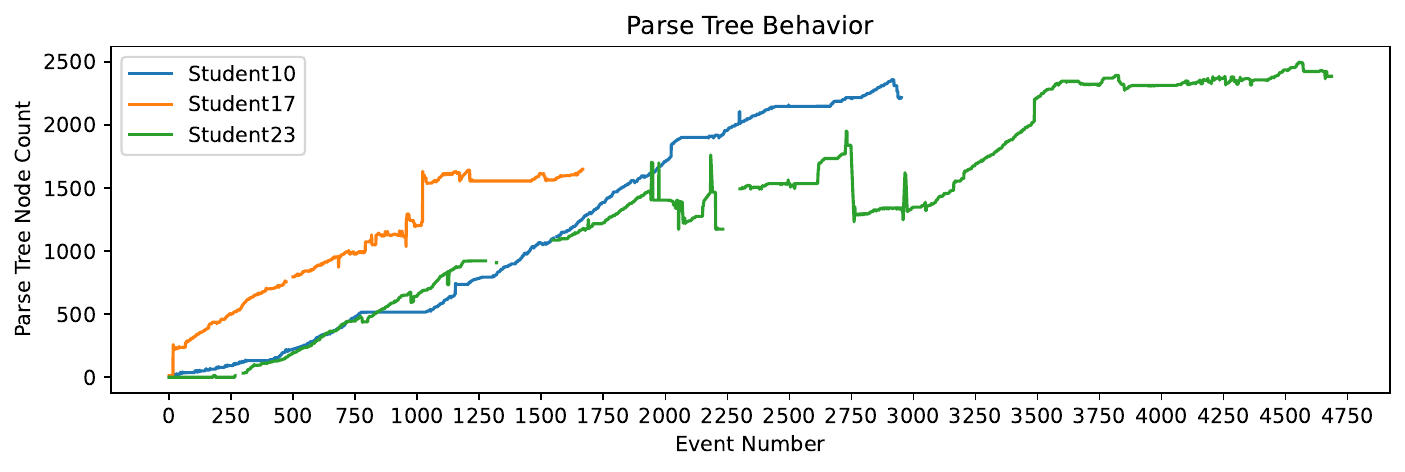}
    \caption{Evolution of parse trees for three students on the same assignment. Each "line" shows the size of the parse tree at each keystroke. Gaps indicate where bridging parse tree construction failed.}
    \label{fig:progress}
\end{figure}

\subsection{RQ3} \label{RQ3}
\begin{figure}
    \begin{minipage}[t]{0.35\textwidth}
        \centering
        \includegraphics[width=1\textwidth]{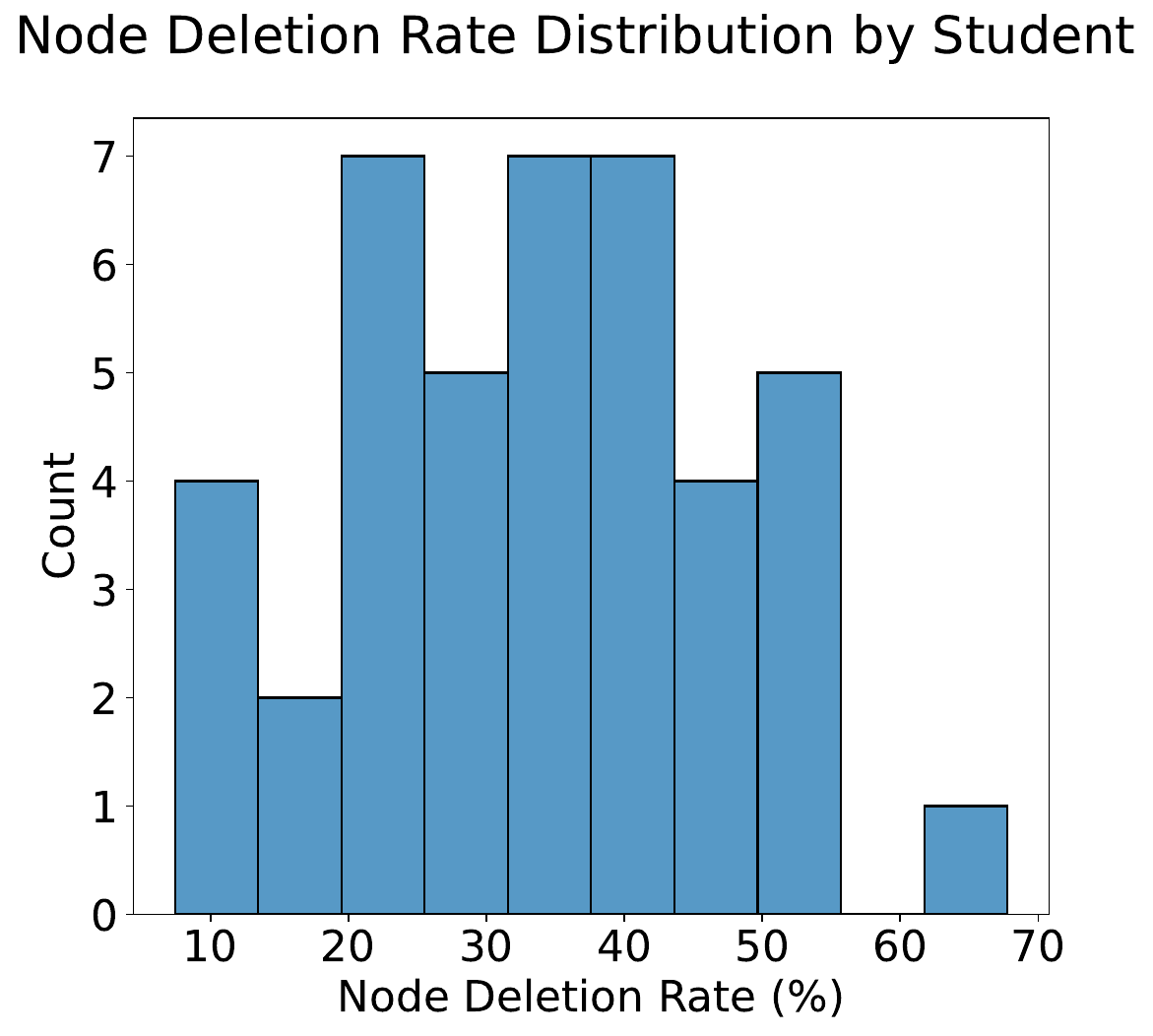}
        \caption{Distribution of students node deletion rate.}
        \label{fig:NDRStudent}
    \end{minipage}
\end{figure}

\begin{figure}
    \centering
    \includegraphics[width=0.45\textwidth]{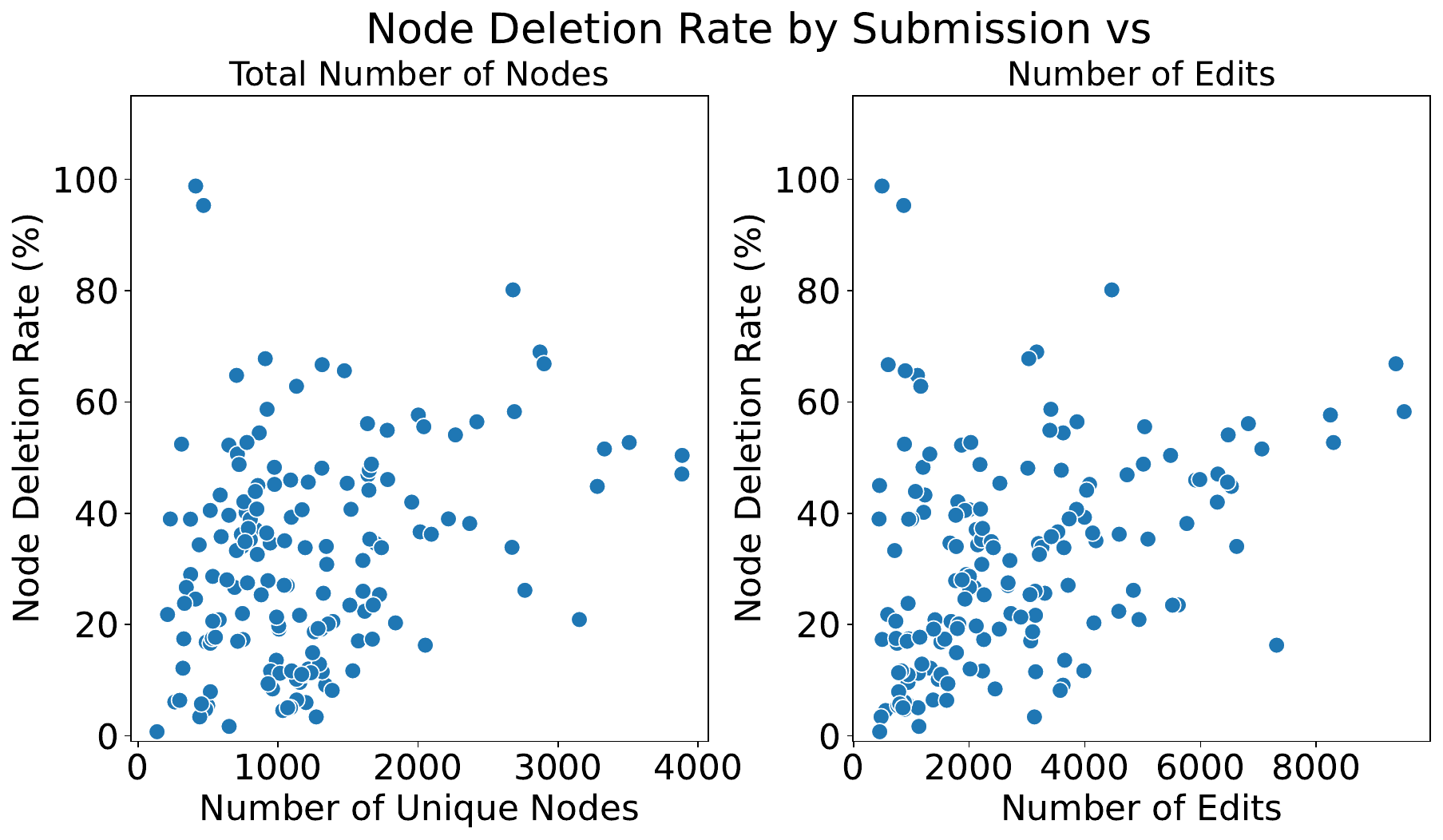}
    \caption{The scatterplot on the left shows the number of unique nodes against node deletion rate. The graph on the right shows the total number of edits vs node deletion rate. Both scatter plots are averaged per student submission.}
    \label{fig:NDRvsNodesSubmission}
\end{figure}
\begin{figure}
\begin{minipage}[t]{0.38\textwidth}
    \centering
\includegraphics[width=0.95\textwidth]{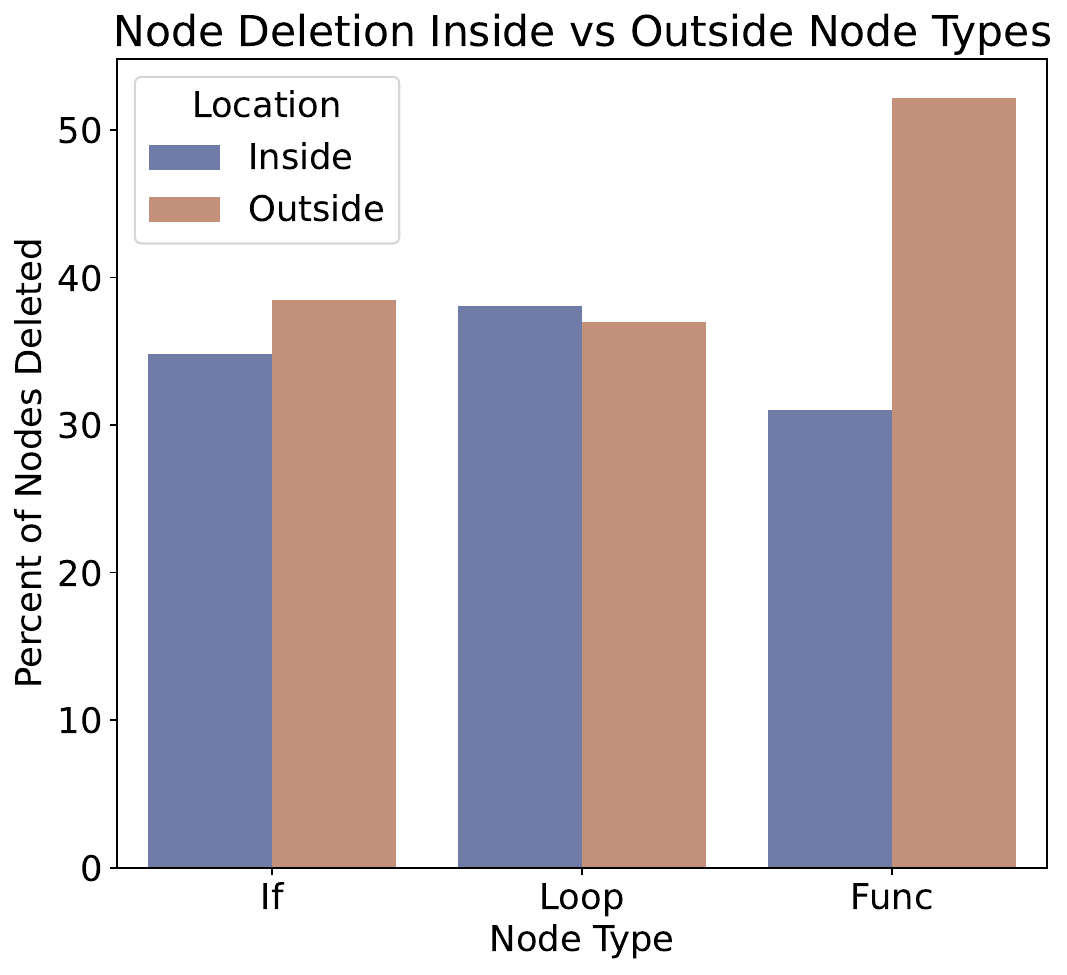}
    \caption{Node deletion rates inside and outside of common programming constructs. A node is inside an if, loop, or function if it is inside the body, or compound statement, of the construct. }
    \label{fig:NDRNodeType}
\end{minipage}
\hspace{0.05\textwidth}
\begin{minipage}[t]{0.38\textwidth}
\centering
\includegraphics[width=0.95\textwidth]{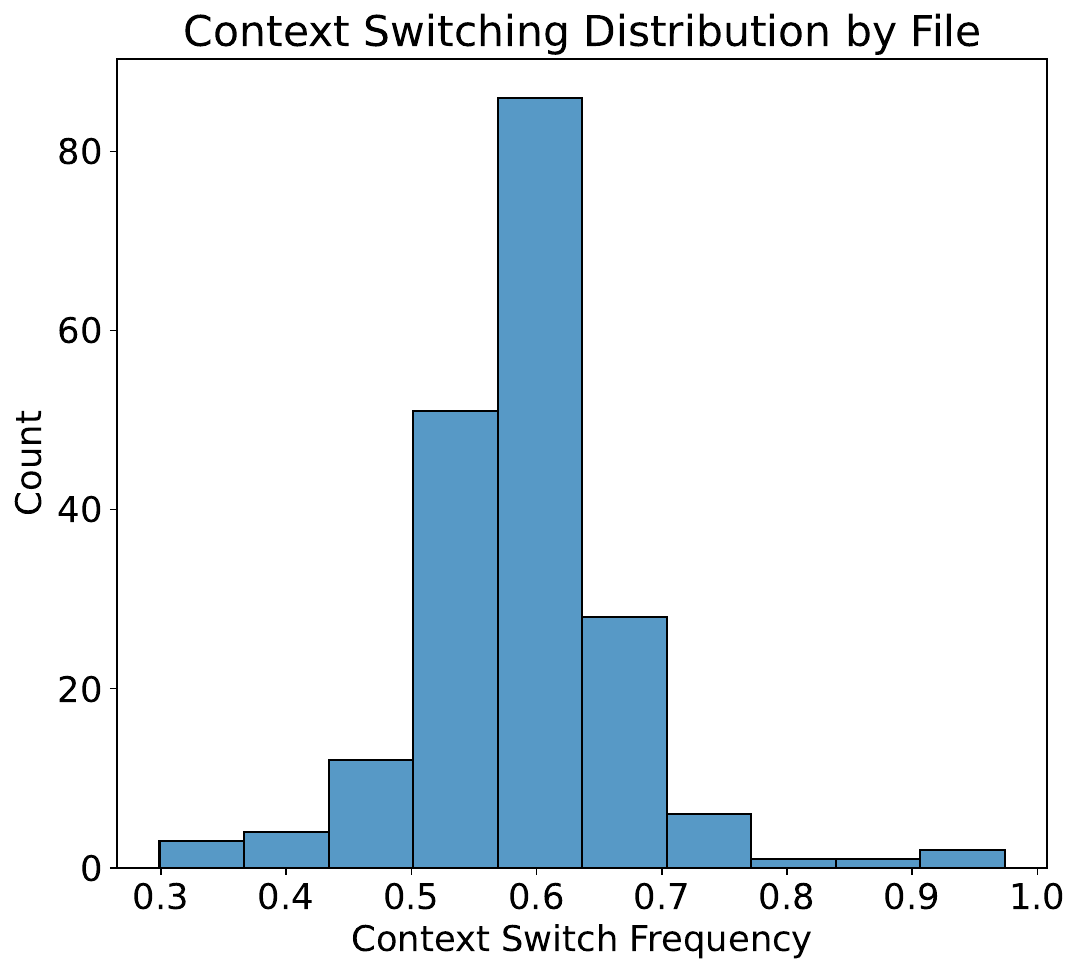}
    \caption{Distribution of context switch frequency averaged per student submission.}
    \label{fig:JumpingDistSubmission}
\end{minipage}
\end{figure}

Our third research question asks, "\rqthree" Deletion can be a useful metric for understanding a number of things about a student. It can potentially show when a student is struggling, or what type of programming concepts they have a harder time with. Brown et al. \cite{brown2024writing} use line deletion as a measure for deletion. Our parse tree node tracking lets us analyze node deletion instead. We suggest that node deletion is a more useful measurement of deletion, as it will note any character deletion that causes some structural change to the code. It also allows for automated information about the deletion, such as what kind of node is being deleted, and what kind of nodes it is a descendant of. 

Deletion of nodes occurred in every file. The median number of occurrences was $41 \ |\ 163$, meaning that in a typical file with over 3,000 keystrokes, the median case contained $163$ deletion edits that resulted in at least one node being removed. On average, $32\%$ of all nodes were eventually deleted ($27 \ |\ 46\%$). Figure \ref{fig:NDRStudent} shows the distribution of each student's average node deletion rate over all assignments. There is a large variation, with some students deleting over half of all nodes they create. The distribution seems like it may be normal, though an expanded dataset with a larger sample size of students would make the distribution more clear. 

Figure \ref{fig:NDRvsNodesSubmission} shows the node deletion rate for each submission against both the total number of nodes, and the number of edits in that submission. There is a positive correlation for both of these comparisons ($r=0.31, p=0.2.9e^{-5}$ and $r=0.32, p=2.2e^{-5}$, respectively). There is a strong correlation between number and edits and number of unique nodes ($r=0.75, p=5.3e^{-32}$), and it would seem that both as students spend longer on their assignment, and create more nodes over time, their rate of deletion increases as well. We note that this is \textit{rate} of node deletion. In other words, students who use more events to complete their assignment tend to delete code more often. This makes sense, as they may be struggling and writing incorrect code that eventually gets deleted. A strong correlation exists between character deletion and node deletion as we expected ($r=0.77, p=1.5e^{-32}$). However, we suggest that node and character deletion rates are not interchangeable, as the node deletion rate measure excludes much of a student's struggle with typing.

By looking at the node deletion rate both inside and outside of a programming construct, we can get an idea if students are overall struggling more with that concept. For example, a deleted \tt{print} statement inside an \tt{if} statement counts as one deletion in an \tt{if} statement. If a string was in the \tt{print} statement, two deletions would be added to the \tt{if}. Figure \ref{fig:NDRNodeType} shows this comparison for three common programming concepts: if statements, for loops, and function definitions. (The function category includes methods inside classes). Two-sample z-tests show that the difference is significant in if statements ($z=-8.9, p<0.0001$), for loops ($z=4.9, p<0.0001$), and functions ($z=-70.2, p<0.0001$). However, despite being different, Cohen's h shows a nearly negligible effect size for conditionals and loops ($h = 0.05, 0.02$ respectively). Contrary to our hypothesis, students appear to struggle about the same whether writing code inside or outside of conditionals and loops. This has a strong implication that students work on code locally, with little awareness of context. For example, if a student is writing a calculation, they will write it similarly whether it is in a loop or not. 

There is a medium effect size for functions ($h = 0.50$), with more code deleted outside of functions than inside -- the opposite of conditionals and loops. This seems counter-intuitive. However, in the dataset we used, with the exception of one assignment, the students were instructed to write their code in functions, and thus the vast majority of code is written in functions or methods. Much of the code written outside of functions was debugging or experimentation code, which would be expected to be deleted before submission.

\subsection{RQ4}  \label{RQ4}

\begin{figure}
    \begin{minipage}[t]{0.3\textwidth}
        \centering
    \includegraphics[width=\columnwidth]{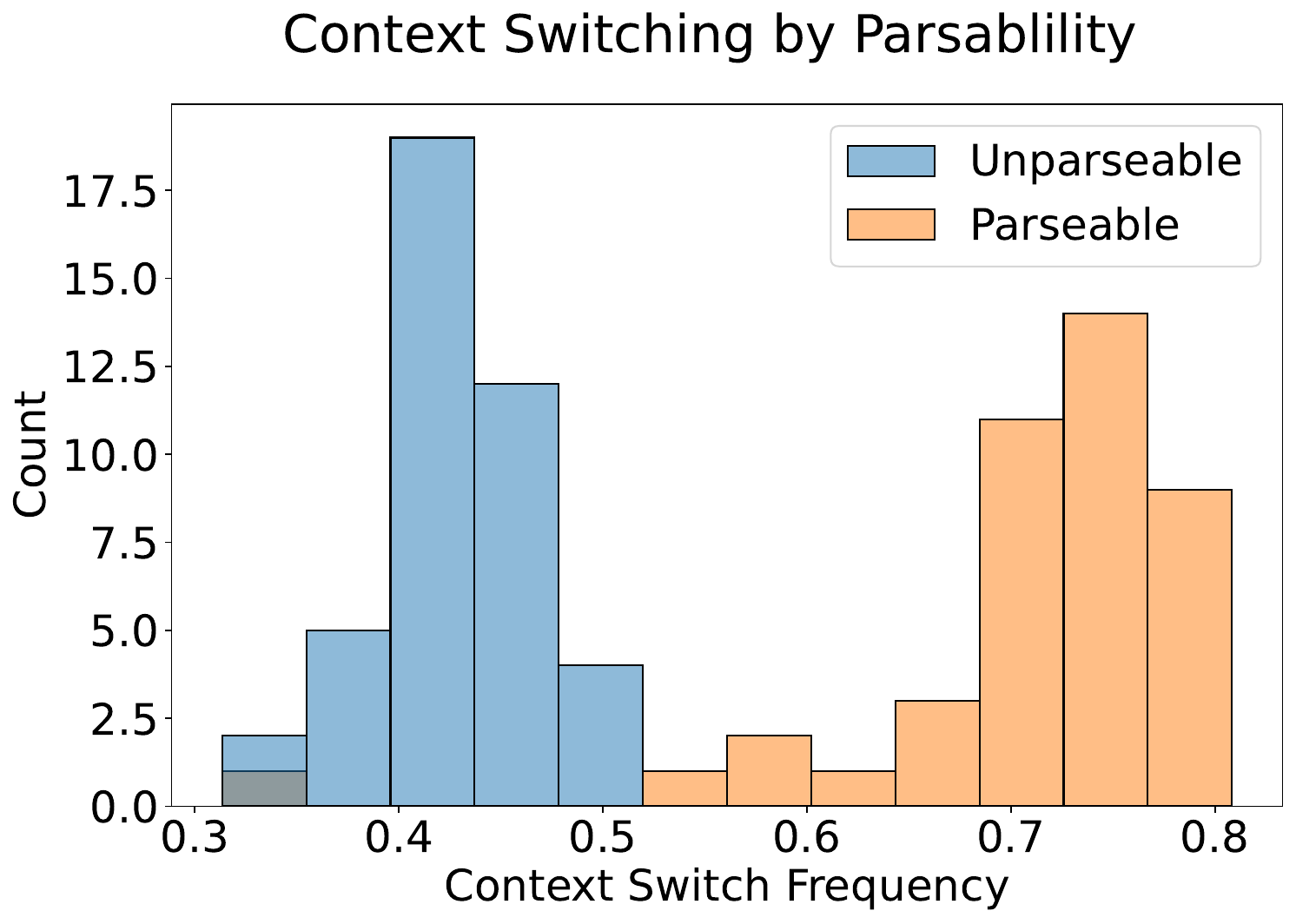}
        \caption{Distribution of the context switch frequency by student submission in parseable and unparseable states.}
        \vspace{-.5\baselineskip}
        \label{fig:JumpingComp}
    \end{minipage}
    \hspace{0.03\textwidth}
    \begin{minipage}[t]{0.425\textwidth}
        \centering
        \includegraphics[width=\columnwidth]{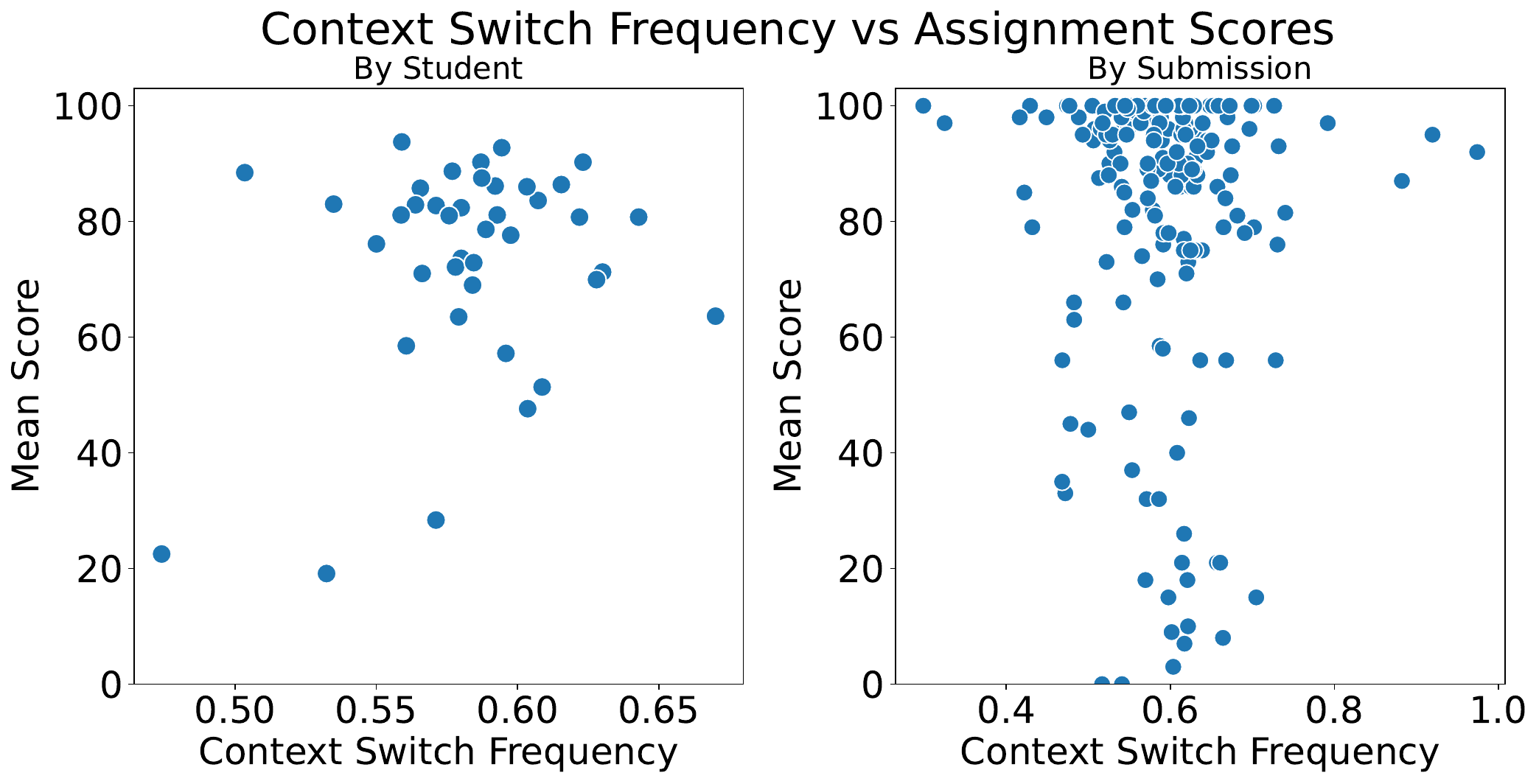}
            \caption{Students' context switch frequency compared to their grades, shown per student on the left, and per submission on the right.}
            \label{fig:JumpvsScoreSubmission}
    \end{minipage}
\end{figure}





Our last research question was "\rqfour" To answer this question, we used context switch frequency which we defined in section \ref{JumpingMetric}. In Figure \ref{fig:JumpingDistSubmission}, we see the distribution of context switch frequency over all submissions. The mean context switch frequency is $58.8\%$, meaning that, on average, $58.8\%$ of the time that students move to working on a different leaf node, the path from the previously edited node is 5 or more edges long. Context switch frequency appears to be normally distributed.

We had hypothesized that context switching would be more common when students are in an unparseable state, as they may be jumping around more to find the cause of the problem. As shown in Figure \ref{fig:JumpingComp}, the opposite was true. Context switching was significantly less frequent when students were in unparseable states ($t=19.5, p=1.6e^{-32}$).  This would imply that when in an unparseable state, students usually aren't flailing to find the location of the error, probably because the line number is given for a compilation error. Furthermore, unparseable states are also not always caused by student errors, but a natural result of writing something like a loop sequentially, so much of the time spent in an unparseable state isn't due to a student's syntax errors. For example, partway through the definition of a for loop, the code will be in an unparseable state, and will stay in an unparseable state until the loop has been fully defined. Most of the time, students will complete this entire sequence mostly linearly, decreasing the context switch frequency. 

Figure \ref{fig:JumpvsScoreSubmission} shows context switch frequency plotted against the achievement metric of assignment grades. These are averaged both by student and by submission. We had hypothesized that there would be a negative correlation, meaning that jumping around a lot would indicate a student who is struggling. This was our only hypothesis test of a programming process feature's relation to an academic outcome because we believed the effect would be obvious, which turned out not to be true. No correlation was found by submission ($r=0.03$, $p=0.71$), but there was a possible small positive correlation when aggregated by student ($r=0.36$, $p=0.03$). This was a surprising result, and could counterintuitively suggest that students who are jumping around less may be struggling more. More analysis is needed to confirm the significance of this finding, but one possible reasoning for this could be that students who are struggling more may spend more time debugging errors in small sections of code. 

We also tested context switch frequency against the number of edits in a submission. We had hypothesized that when students spend longer on an assignment, they might be struggling more, and thus jumping around in their code at a higher rate. This wasn't supported by our analysis, as there was no statistically detectable correlation ($r=-0.07$, $p=0.33$).

\subsection{Threats to validity}
\subsubsection{Algorithms:}\ These algorithms have primarily been tested on one publicly available dataset of keystrokes from students in a university-level introductory programming course  \cite{DVN/BVOF7S_2022, edwards2023review}. To ensure that our algorithms are robust and general, testing and validation on a larger variety of data is required. We call on researchers producing keystroke data to make their datasets publicly available to enable better external validation of studies like ours.

In this paper, we use a definition for a bridging parse tree that attempts to represent what unparseable code will become. We suggest that this is the most useful definition for analysis, but there are other possible definitions of what constitutes a "correct" tree for unparseable code. For example, in Figure \ref{fig:BridgingTree}, there is an \textit{Arguments} node in state $t+1$. One could argue that there shouldn't be an instance of some node type in a parse tree if it is missing one of that node type's semantically required children, such as the \textit{Arguments} node missing its \textit{)} child. 

\subsubsection{Analysis:}\ Given that we only included files with at least $80\%$ successful parse trees, many files had to be omitted from analysis. Some students or submissions may be inaccurately portrayed by the subset of files that were able to be included. See Section \ref{BridgeIssues} for analysis tempering this threat.
Lastly, our analysis only encompassed one dataset of keystroke data. Any of our findings may not prove to hold for a larger, more varied population.

\section{Conclusions}
While static analysis of code quality and structure is a mature research stream with a rich literature corpus, analysis of the process programmers use to write code, on the other hand, has not received nearly as much attention. Possible reasons for this could include lack of datasets and/or lack of analysis tools. Our work presented in this paper does not make any additional data available, but it does present tools and techniques that enable analysis at scale. Our work also presents results that have broad implications in our understanding of how students write code. We suggest that it could contribute to increased awareness of, and interest in, programming process as an important and impactful research discipline.

The first half of our paper focused on algorithms for parse tree node tracking through time. Our algorithms unlock our ability to quantitatively measure programming behaviors using much larger datasets than before. These behaviors include amount of time spent working on particular syntactic constructs, commenting and uncommenting code, writing code top-down or bottom-up, renaming variables, shifting focus between different parts of code, deletion of characters or blocks of code, and many other behaviors. The results of our analysis presented in this paper are potentially impactful on their own, but they also highlight the potential of syntactic construct tracking through time as an automatic analysis tool, relieving investigators of the significant manual work required previously~\cite{brown2024writing}.

The second half of our paper focused on analysis of a public keystroke dataset. We found, counter to our expectations, that students delete code at similar rates whether inside or outside of conditionals and loops, a large amount of commented out code is eventually restored, and that the amount of jumping around students do in code is not necessarily correlated with academic outcomes. Also to our surprise, our replication study yielded, in some cases, vastly different results from~\citet{brown2024writing}, including frequency of exterior pastes, renaming variables, commenting and uncommenting code, and writing comments. We suggest that these differences say little about the internal validity of the studies but have significant implications to external validity. The contexts of the two datasets are very different, resulting in strikingly varied programming behaviors, reinforcing the importance of considerations about generalizability in programming process studies. Having automated analysis tools, such as those presented in this paper, will take on increasing importance as studies utilize more datasets from different contexts.

Looking forward, automated analysis tools will become more important for another reason. Generative AI-assisted and agent-based programming are quickly changing the way students and professionals approach writing a computer program. Discoveries of how students write programs likely will have shorter relevance lifetimes than ever before, and so automated tools will be needed to keep our understanding current without the analyses requiring excessive manual labor. And an up-to-date view of these processes will be important as educators grapple with how to incorporate AI-based tools into CS pedagogy and curriculum.

\bibliographystyle{ACM-Reference-Format}
\balance
\bibliography{paper}

\end{document}